\documentclass[a4paper,fleqn,usenatbib]{mnras}

\usepackage[T1]{fontenc}
\usepackage{ae,aecompl}

\usepackage{graphicx}	% Including figure files
\usepackage{amsmath}	% Advanced maths commands
\usepackage{amssymb}	% Extra maths symbols

\title[An Off-Axis Galaxy Cluster Merger: Abell 0141]{An Off-Axis Galaxy Cluster Merger: Abell 0141}

\author[Turgay CAGLAR]{
Turgay CAGLAR\thanks{E-mail: caglar@strw.leidenuniv.nl}
\\
Leiden Observatory, Leiden University, PO Box 9513, 2300 RA Leiden, The Netherlands\\
Y{\i}ld{\i}z Technical University, Faculty of Science and Art, Department of Physics, Istanbul 34220, Turkey\\
}
\date{Accepted XXX. Received YYY; in original form ZZZ}

\pubyear{2018}
   \volume{475}

\begin{document}
\label{firstpage}
\pagerange{2870-2877}
\maketitle

\begin{abstract}
We present structural analysis results of Abell 0141 (\textit{z} = 0.23) based on X-ray data. The X-ray luminosity map demonstrates that Abell 0141 (A0141) is a bimodal galaxy cluster, which is separated on the sky by $\sim$ 0.65 Mpc with an elongation along the north-south direction. The optical galaxy density map also demonstrates this bimodality. We estimate sub-cluster ICM temperatures of 5.17$^{+0.20}_{-0.19}$ keV for A0141N and 5.23$^{+0.24}_{-0.23}$  keV for A0141S. We obtain X-ray morphological parameters w = 0.034$\pm{0.004}$, c = 0.113$\pm{0.004}$ and w = 0.039$\pm{0.004}$, c = 0.104$\pm{0.005}$ for A0141N and A0141S, respectively. The resulting X-ray morphological parameters indicate that both sub-clusters are moderately disturbed non-cool core structures. We find a slight brightness jump in the bridge region, and yet, there is still an absence of strong X-ray emitting gas between sub-clusters. We discover a significantly hotspot ($\sim$ 10 keV) between sub-clusters, and a Mach number $\textit{M}$ = 1.69$^{+0.40}_{-0.37}$ is obtained by using the temperature jump condition. However, we did not find direct evidence for shock-heating between sub-clusters. We estimate the sub-clusters' central entropies as $K_{0}$ > 100 keV cm$^{2}$, which indicates that the sub-clusters are not cool cores. We find some evidence that the system undergoes an off-axis collision; however, the cores of each sub-clusters have not yet been destroyed. Due to the orientation of X-ray tails of sub-clusters, we suggest that the northern sub-cluster moves through the south-west direction, and the southern cluster moves through the north-east direction. In conclusion, we are witnessing an earlier phase of close core passage between sub-clusters.
\end{abstract}


%%%%%%%%%%%%%%%%%%%%%%%%%%%%%%%%%%%%%%%%%%%%%%%%%%

%%%%%%%%%%%%%%%%% BODY OF PAPER %%%%%%%%%%%%%%%%%%

%%%%%%%%%%%%%%%%%%%%%%%%%%%%%%%%%%%%%%%%%%%%%%%%%%
%%%%%%%%%%%%%%%%%%%%%%%%%%%%%%%%%%%%%%%%%%%%%%%%%%
%%%%%%%%%%%%%			INTRODUCTION 			%%%%%%%%%%%%%%%%
%%%%%%%%%%%%%%%%%%%%%%%%%%%%%%%%%%%%%%%%%%%%%%%%%%
%%%%%%%%%%%%%%%%%%%%%%%%%%%%%%%%%%%%%%%%%%%%%%%%%%

\section{Introduction}
Clusters of galaxies are the largest observed laboratories, which are formed around massive galaxies due to gravitational in-fall of other objects. Ongoing studies of these laboratories allow us to understand structure formation of the universe. X-ray investigations of clusters of galaxies have demonstrated the existence of hot plasma, which surrounds member galaxies in the vicinity, namely intra-cluster medium (ICM).  Another interesting result of these studies is the presence of massive bounding objects, such as cluster-group and cluster-cluster interactions, which are known as mergers.

The essential concept of cluster mergers is energy and angular momentum transfer between gas clumps, which surround galaxy cluster environments. The dynamical process throughout merger events can cause ram pressure stripping, turbulence, shock fronts, and cold fronts in the cluster vicinity \citep[e.g.,][]{Markevitch2001,Markevitch2007}. Another important sign of merger interactions is positional offset between X-ray and optical centroids of bright central galaxies (BCGs) \citep[e.g.,][]{Smith2005,Shan2010}. Previous studies of merging galaxy clusters have demonstrated that physical process can be separated to three topics: early(pre), ongoing and late(post) stage cluster mergers. There are a lot of examples for ongoing and late mergers in the literature \citep[e.g.,][]{Briel1991,Roettiger1995,Markevitch1999,Markevitch2002,Kempner2005,Russell2010,Caglar}; however, early stages of mergers are still quite a few \citep{Fujita1996,Werner,Ogrean2015,Akamatsu2016,Bulbul2016}. Consequently, the physical mechanisms of early  stage mergers are still not well-understood, and observations of new objects are critical to understand the mechanism. 

Here, we present a multi-wavelength investigation of a merging galaxy cluster based on X-ray analysis results: Abell 0141 (A0141). Since early stages of mergers are still poorly understood, we aim to understand the physical mechanisms of A0141 using X-ray morphology and temperature distributions. Therefore, X-ray analysis is performed to explain the physical process through merging events.

Abell 0141 is a BM III type distant galaxy cluster with an Abell distance D = 6 \citep{Bautz} and redshift \textit{z} = 0.23 \citep{Struble1,Struble2}. Due to its morphological type, the cluster is dominated by four bright elliptical galaxies rather than a central dominant galaxy. The brightest galaxy is 0.4 mag brighter than the second brightest one \citep{Krick}. A0141 has a galaxy distribution, which is roughly elongated through the north and south direction \citep{Dahle}. X-ray analysis of A0141 was performed with \textit{ROSAT} data by \citet{Ebeling1996} resulting in kT = 9.2 keV and L$_{x}$ = 1.3$\times$10$^{45}$ erg s$^{-1}$. At that time, the cluster's average temperature was reported 8.0 keV (\textit{ROSAT} data), 5.31 keV (\textit{Chandra} data) by the following studies \citet{Cruddace} and \citet{Cavagnolo}, respectively. \citet{Piffaretti} used \textit{ROSAT} data to measure cluster's mass and luminosity at the overdensity radius of 500 and provided the following results: r$_{500}$ = 1096 kpc, L$_{500}$ = 5.16$\times$10$^{44}$ erg s$^{-1}$, and M$_{500}$ = 4.72$\times$10$^{14}$ M$_{\odot}$. Furthermore, \citet{Comis} studied \textit{Chandra} data and reported a gas mass M$_{gas}$ = 3.4$\times$10$^{13}$ M$_{\odot}$ and a total mass M$_{total}$ = 5.8$\times$10$^{14}$ M$_{\odot}$ within an over-density radius of 2500. On the other hand, \citet{Martino} surprisingly estimated a gas mass M$_{gas}$ = 0.08$\times$10$^{13}$ M$_{\odot}$ and a total mass M$_{total}$ = 0.8$\times$10$^{14}$ M$_{\odot}$ within an over-density radius of 2500. Finally, \citet{Okabe} determined cluster's mass M$_{Vir}$ = 5.67$\times$10$^{14}$ M$_{\odot}$ with weak-lensing analysis.

The paper is organized as follows: in section 2, we present the observation logs and data processing. In section 3, we describe X-ray analysis procedures. Section 4 explains structural analysis procedures of A0141.  In section 5, we discuss our results. Finally, we summarize our results in section 6. Throughout this paper, we adopt a standard $\Lambda$ cold dark matter cosmology parameters: H$_{0}$ = 70 km s$^{-1}$ Mpc$^{-1}$, $\Omega$$_{M}$ = 0.3 and $\Omega$$_{\Lambda}$ = 0.7 for a flat universe. In this cosmology, 1$\arcmin$ corresponds to 219.79 kpc. Unless stated otherwise, the error values are quoted at the 90\% confidence interval in our analysis. 

%%%%%%%%%%%%%%%%%%%%%%%%%%%%%%%%%%%%%%%%%%%%%%%%%%
%%%%%%%%%%%%%%%%%%%%%%%%%%%%%%%%%%%%%%%%%%%%%%%%%%
%%%%%%%%%%%%    OBSERVATION AND DATA PROCESSING    %%%%%%%%%%%%%%
%%%%%%%%%%%%%%%%%%%%%%%%%%%%%%%%%%%%%%%%%%%%%%%%%%
%%%%%%%%%%%%%%%%%%%%%%%%%%%%%%%%%%%%%%%%%%%%%%%%%%

\section{Observations and Data Processing}
An \textit{XMM-Newton} observation was performed on 2012 June 12 for exposure of 31.1 ks with the observation id 0693010501. A thin filter was used for both MOSs and pn cameras. X-ray observations were taken in full frame for MOSs and extended full frame for pn. A Chandra observation of A0141 was performed in VFAINT mode on 2008 August 11 for 19.9 ks with observation id 9410.  X-ray observational data were gathered from \textit{XMM-Newton} Science Archive (XSA) and \textit{Chandra} Data Archive. We performed data reduction using High Energy Astrophysics Software \textit{HEAsoft v6.22}, \textit{XMM-Newton} Science Analysis Software (\textit{XMM-SAS v16.1}) and \textit{XMM-Newton} Extended Source Analysis Software (\textit{XMM-ESAS}). The \textit{Chandra} data was reprocessed with Chandra Interactive Analysis of Observations Software \textit{ciao-4.9} and Current Calibration Database \textit{CALDB-4.7.5.1}. X-ray observation data logs are listed in Table \ref{obslog}.

Current calibration (ccf) and summarized observation data files (odf) are generated using the tasks: cifbuild-4.8 and odfingest-3.30, respectively. The emchain-11.19 and epchain-8.76.1 tasks were applied to data, which generates MOS and pn event files, respectively. The light curve was created to determine a final net data, and corrupted data was extracted performing evselect-3.62. Finally, point-like X-ray sources within the galaxy cluster are detected using edetect\_chain-3.15 and removed from the data in our analysis. The level 2 event file was reprocessed with ciao task chandra\_repro, which also generates a mask and a bad pixel file.

\begin{table}
\begin{center}
\caption{Log of X-ray Observations.}
\begin{tabular}{@{}cccl@{}} 
\hline
\hline
ObsID	    	& 	Satellite		& 	Date Obs		& Effective Exposure   	\\
            		&                   		&               		& Time (ks)          	 	\\
\hline
09410       	& \textit{Chandra}  	& 2008-08-11    	& ACIS 19.9 (\%92.5)       	\\
0693010501  	&\textit{XMM-Newton}& 2010-06-12    	& MOS1 30.9 (\%75.2)	\\
			& 				& 				& MOS2 30.9 (\%82.1)	\\
			& 				& 				& PN 27.0 (\%41.3)	     	\\
\hline
\end{tabular}
\label{obslog}
\end{center}
\end{table}

%%%%%%%%%%%%%%%%%%%%%%%%%%%%%%%%%%%%%%%%%%%%%%%%%%
%%%%%%%%%%%%%%%%%%%%%%%%%%%%%%%%%%%%%%%%%%%%%%%%%%
%%%%%%%%%%				 ANALYSIS	       %%%%%%%%%%%%%%%%%%%%%
%%%%%%%%%%%%%%%%%%%%%%%%%%%%%%%%%%%%%%%%%%%%%%%%%%
%%%%%%%%%%%%%%%%%%%%%%%%%%%%%%%%%%%%%%%%%%%%%%%%%%

\section{Analysis}

\begin{figure*}
 \begin{center}
  \includegraphics*[width=7.7cm]{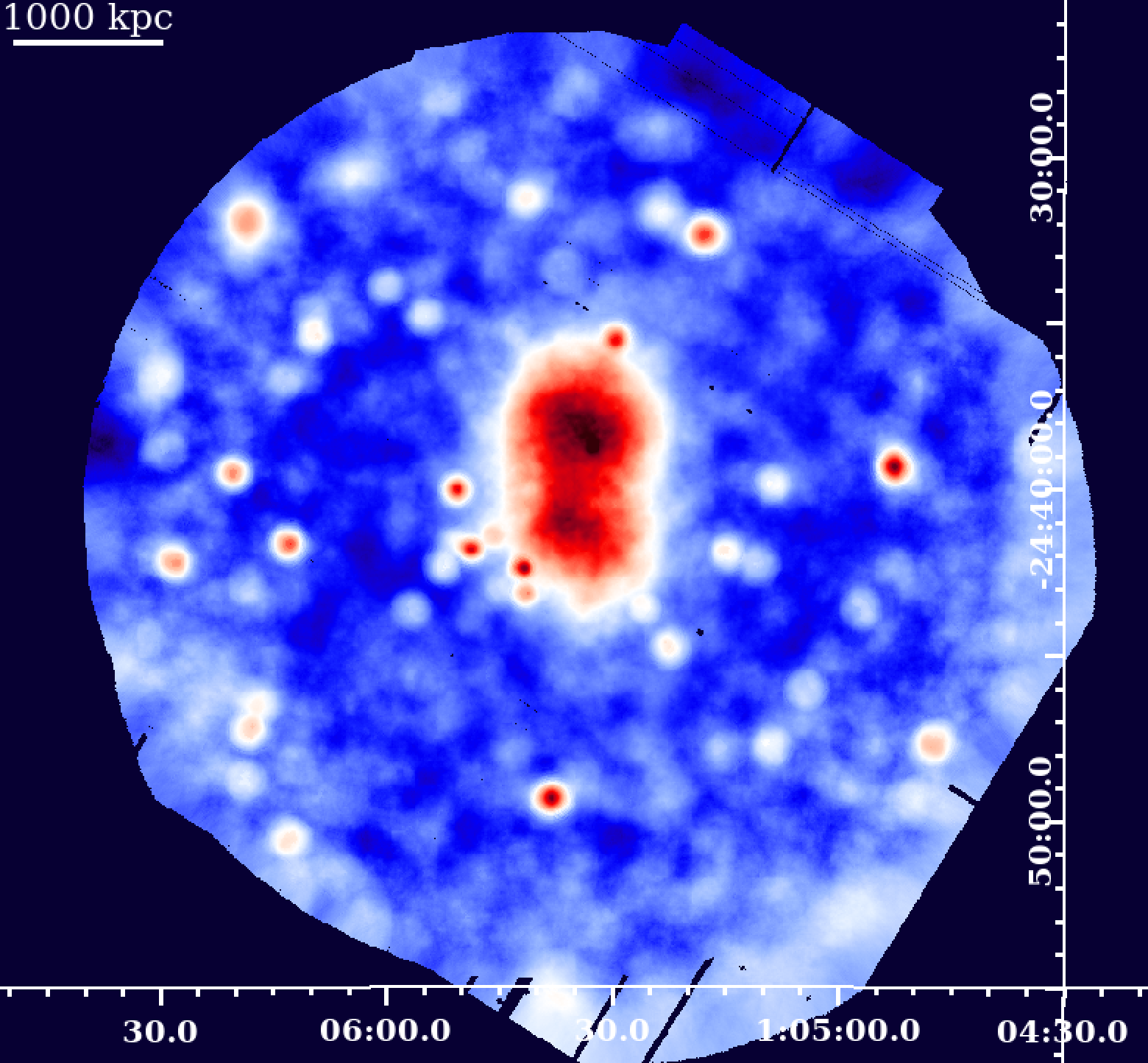}
   \includegraphics*[width=8.35cm]{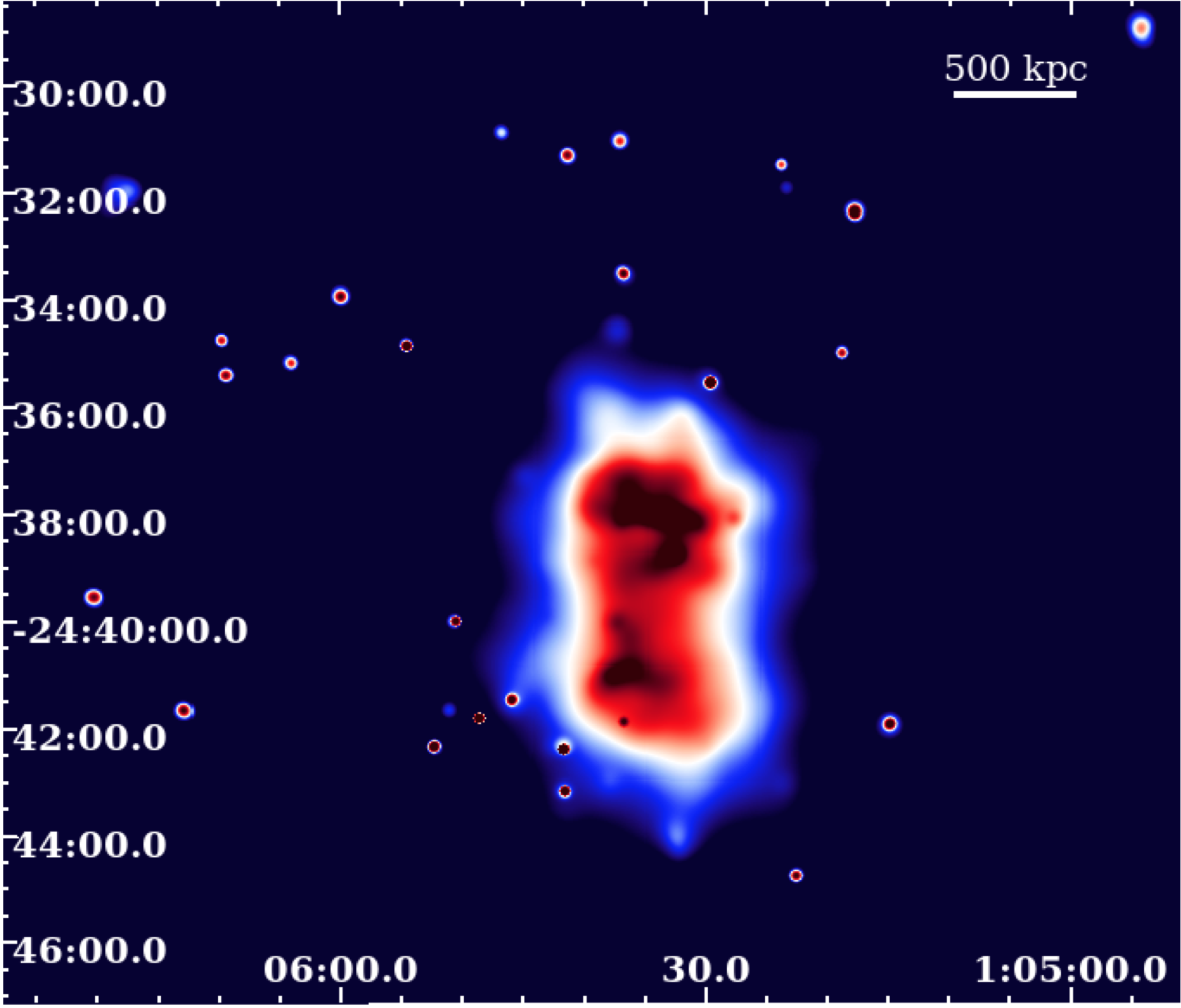}
 \caption{\label{back-sub-im} The background subtracted, exposure- and vignetting-corrected, and adaptively-smoothed X-ray images. \textbf{Left:} \textit{XMM-Newton} \textbf{Right:} \textit{Chandra} The connecting region between sub-clusters can directly be seen from the images.}
\end{center} \end{figure*}

\subsection{Spatial Analysis}
The background subtracted, exposure- and vignetting corrected, combined \textit{XMM-Newton} X-ray image is generated using the following analysis procedure. The MOS and pn raw data are generated by the mos-filter and the pn-filter task, respectively. The \textit{XMM-Newton} X-ray images are created in the energy range of 0.4-10.0 keV for MOS and pn using the mos-spectra and the pn-spectra, respectively. The point-like X-ray sources within A0141 are detected using the cheese task above a flux threshold 5 $\times$ 10$^{-15}$ erg cm$^{-2}$ s$^{-1}$. The detected point sources are removed by the mask parameter in the analysis. Since the \textit{XMM-ESAS} performs analysis with detector coordinates, the rot-im-det-sky task is used to transform detector coordinates into sky coordinates. The soft proton and background contaminations are excluded from the data using the spectral parameters by proton\_scale and proton tasks, respectively. Finally, MOSs and pn data are combined with the comb task, and the combined image is smoothed adaptively by the adapt task. We remind you that the \textit{XMM-Newton} packages from Extended Source Analysis Software version 0.9.40 are used for the analysis procedure of image generation. The exposure-corrected \textit{Chandra} image is generated using fluximage, and point-like sources are detected using \textit{ciao} task wavdetect. A background subtracted image is generated using blanksky\_image for \textit{Chandra} data. The background subtracted, exposure- and vignetting-corrected, adaptively-smoothed \textit{XMM-Newton} and the \textit{Chandra} images are presented in Fig. \ref{back-sub-im}. 

The surface brightness is defined as the projected plasma emissivity per area on the sky. The X-ray surface brightness of each sub-cluster is fitted with a $\beta$ model \citep{Cavaliere}. The model is defined as: 

\begin{equation}
S (r) = S_{0} \times \left[ 1 + \left(r/r_{c}\right)^{2} \right]^{-3\beta+0.5} + c
\end{equation}
where S$_{0}$ is the central surface brightness, r$_{c}$ is the core radius, and $\beta$ is the shape parameter. The background level (c) is crucial to estimate best-fitting parameters of the surface brightness, and the background level is found to be 2.22$\pm{0.22}$ $\times$ 10$^{-2}$ counts arcsec$^{-2}$. The resulting parameters of $\beta$ model are presented in Table \ref{surbri} and are further discussed in section 5.1. 

\subsection{Spectral Analysis}
For the spectral analysis of \textit{XMM-Newton}, we used the evselect-3.62 task to generate spectrum and background files. The response files of A0141 were created using the rmfgen-2.2.1 and arfgen-1.92 tasks. The local background was subtracted from the source file using an annular region within 11$\arcmin$ - 12$\arcmin$ away from cluster's centre. The instrumental background lines: Al K-$\alpha$, Si K-$\alpha$ and Cu K-$\alpha$ are carefully removed from the data. All generated spectral files were grouped by grppha. \textit{Chandra} spectral data were processed with specextract, and the blank sky background files were generated using \textit{ciao} task blanksky. We perform the spectral analysis of A0141 using XSPEC-12.9.1m. The thermal model APEC was used with photoelectric absorption model tbabs in our analysis to estimate plasma temperature. The data is fitted within the energy range of 0.3 - 10.0 keV for \textit{XMM-Newton} and 0.5 - 7.0 keV for \textit{Chandra}. \citet{Lodd} abundance table was applied in the plasma emission and photoelectric absorption models. The spectral fit results of A0141 are presented in Tables \ref{spect} and \ref{thermal}.

\section{The Structural Analysis of A0141}

\subsection{Temperature Structure}
\begin{figure}
 \begin{center}
  \includegraphics*[width=7.6cm]{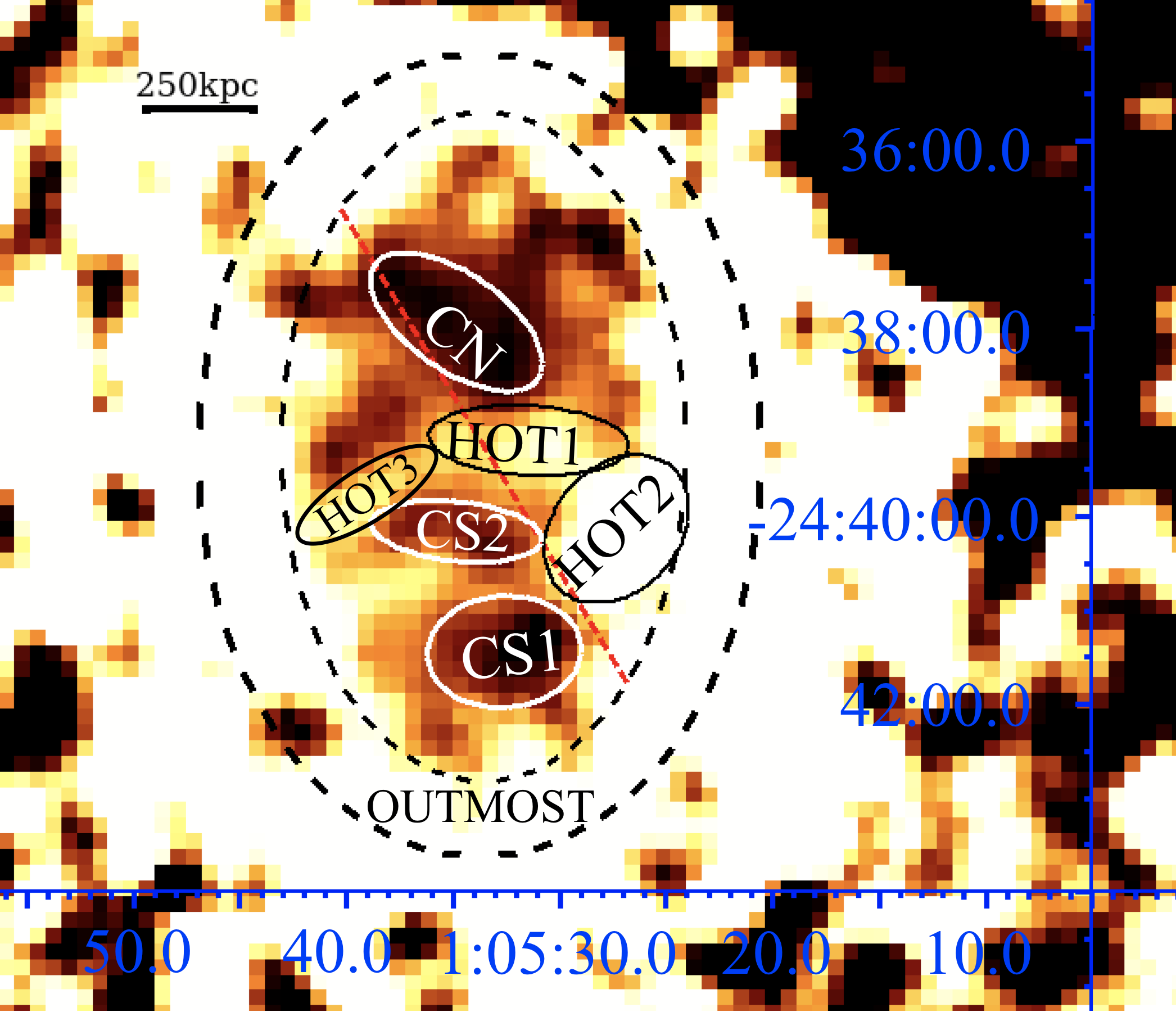}
 \caption{\label{tempmap} Temperature map of A0141.}
\end{center} 
\end{figure}

The temperature map of A0141 is generated by using the hardness ratio approximation.  The hardness ratio map is obtained by dividing hard and soft images. The energy range of soft and hard images are selected within 0.7 - 1.5 keV and 1.6 - 7.0 keV, respectively. In these energy ranges, the hard X-ray photon counts are equal to the soft X-ray photon counts. To avoid non-ICM effects, we did not include the super-soft band data (< 0.7 keV) due to galactic absorption and the hard band data (> 7.0 keV) due to active galactic nucleus contaminations. The point sources are removed from the data to prevent soft proton contamination. The point source gaps are filled with \textit{ciao} task \textit{dmfilth}. A theoretical conversion factor is applied to the hardness ratio map to obtain the temperature map. Similar methods are outlined in different galaxy clusters \citep[e.g.,][]{Ferrari,Caglar}. In the colour-coding of temperature map, the black roughly corresponds to 4.0 - 5.5 keV, the brown roughly corresponds to 5.5 - 7.0 keV, and the yellow roughly corresponds to 7 - 10.0 keV. The outer regions of our sample are affected from low count statistics, which give false information. The temperature map of A0141 is presented in Fig. \ref{tempmap}, which is obtained from XMM-Newton data. Due to the temperature map, A0141 is characterized by two cold sub-structure cores ($\sim$ 4 - 6 keV) and a hot region (9.84 keV) between them. However, we note that calculation of systematic errors is critical for the spectral analysis in low surface brightness regions. The systematic errors can be caused by the following effects: (i) X-ray background contamination, (ii) PSF contamination, and (iii) unresolved point source contamination. In our survey, the hot region between sub-clusters is affected by low count statistics, since there is no strong X-ray ICM. Therefore, we added the systematic errors to the statistical errors within 90\% confidence. A Mach number ($\textit{M}$ = 1.69$^{+0.40}_{-0.37}$) is estimated for the hot region and the sound speed in front of the shock is c$_{S}$ = 1231$^{+100}_{-89}$ km s$^{-1}$. By multiplying Mach number and sound speed $\textit{M}$ $\times$ c$_{S}$ , we calculate a shock velocity V$_{shock}$ =  2081$^{+671}_{-601}$ km s$^{-1}$. A Mach number ($\textit{M}$ = 1.37$\pm{0.10}$) can also be estimated from the surface brightness density jump, which results in V$_{shock}$ = 1686$^{+663}_{-541}$ km s$^{-1}$.

\begin{table}
\begin{center}
\caption{The spectral best-fitting parameters of A0141, which represents the sub-clusters' means with $^1$\textit{XMM-Newton}  and $^2$\textit{Chandra} data.}
\begin{tabular}{@{}l c c r@{}} 
\hline
\hline
Region			& 	$kT$					& Abundance  				& $\chi^2/dof$ 			\\	
				& 	(keV) 				& ($Z_{\odot}$) 			& 	 		 		\\
\hline
A0141N$^{1}$	 	&	 5.17$^{+0.20}_{-0.19}$		&	0.30$\pm{0.08}$			& 	1062/1075 = 0.99		\\
A0141S$^{1}$ 		& 	5.23$^{+0.24}_{-0.23}$ 		&	0.29$\pm{0.09}$			&	1215/1417 = 0.86		\\
A0141N$^{2}$ 		& 	4.96$^{+0.84}_{-0.66}$  	&  	0.35$^{+0.24}_{-0.22}$	&	203/188 = 1.08		\\
A0141S$^{2}$ 		& 	5.63$^{+0.72}_{-0.70}$  	&	0.39$^{+0.27}_{-0.25}$	&	330/329 = 1.00		\\
\hline
\end{tabular}
\label{spect}
\end{center}
\end{table}

\begin{table}
\begin{center}
\caption{The spectral best-fitting parameters of A0141 with $^1$\textit{XMM-Newton}  and $^2$\textit{Chandra} data.}
\begin{tabular}{@{}l c c r@{}} 
\hline
\hline
Region			& 	$kT$					& Abundance  				& $\chi^2/dof$ 			\\	
				& 	(keV) 				& ($Z_{\odot}$) 			& 	 		 		\\
\hline
CN$^{1}$			 	&	 4.74$^{+0.30}_{-0.29}$	&	0.29$\pm{0.09}$		& 	788/803 = 0.98		\\
CN$^{2}$			 	&	 4.64$^{+0.67}_{-0.52}$	&	0.17$^{+0.19}_{-0.14}$		& 	189/190 = 1.00		\\
CS1$^{1}$ 			& 	5.62$^{+0.59}_{-0.55}$  	&	0.75$^{+0.28}_{-0.26}$	&	307/307 = 1.00		\\
CS1$^{2}$ 			& 	5.45$^{+1.56}_{-1.13}$  	&	0.64$^{+0.58}_{-0.49}$	&	137/149 = 0.92		\\
CS2$^{1}$ 			& 	5.87$^{+1.01}_{-0.81}$  	&  	0.22$^{+0.26}_{-0.21}$	&	131/167 = 0.79		\\
HOT 1$^{1}$			& 	9.84$^{+2.77}_{-1.75}$  	&	0.3 (fix)				&	156/204 = 0.78		\\
HOT 1$^{2}$			& 	10.31$^{+4.43}_{-3.63}$  	&	0.3 (fix)				&	92/109 = 0.85		\\
HOT 2$^{1}$			& 	7.30$^{+2.63}_{-1.62}$  	&	0.3 (fix)				&	96/120 = 0.80		\\
HOT 3$^{1}$			&	8.26$^{+2.13}_{-1.51}$	& 	0.3 (fix)				&	128/158=0.81		\\
OUTMOST$^{1}$		&	5.80$^{+1.72}_{-1.21}$	&	0.3 (fix)				&	402/402=1.00		\\
\hline
\end{tabular}
\label{thermal}
\end{center}
\end{table}

 \begin{figure*}
 \begin{center}
  \includegraphics*[width=7.5cm]{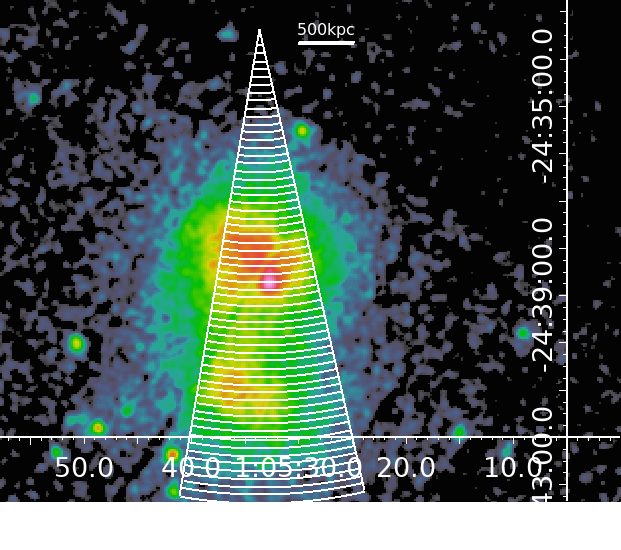}
      \includegraphics*[width=8.9cm]{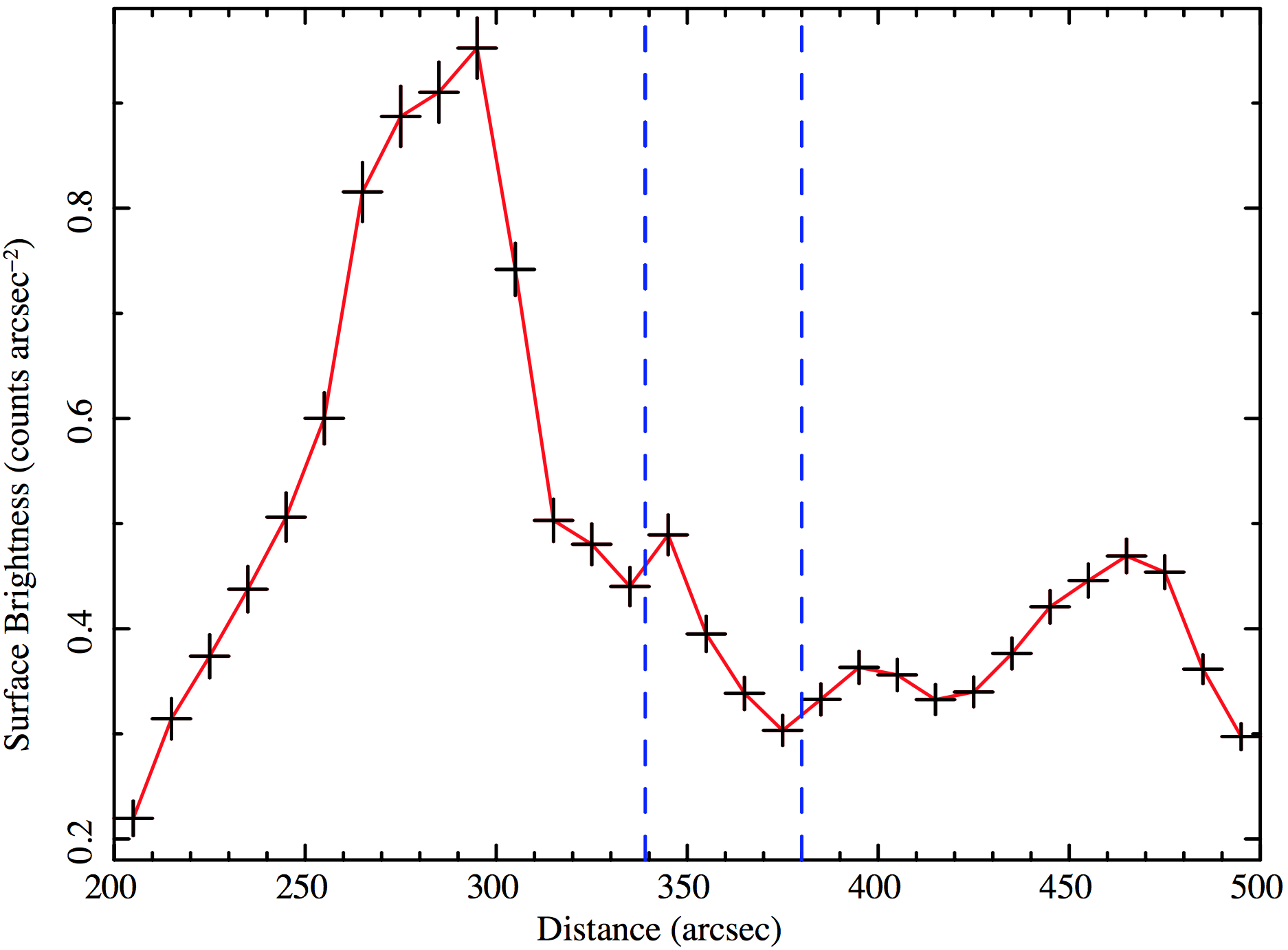}
          \includegraphics*[width=8.5cm]{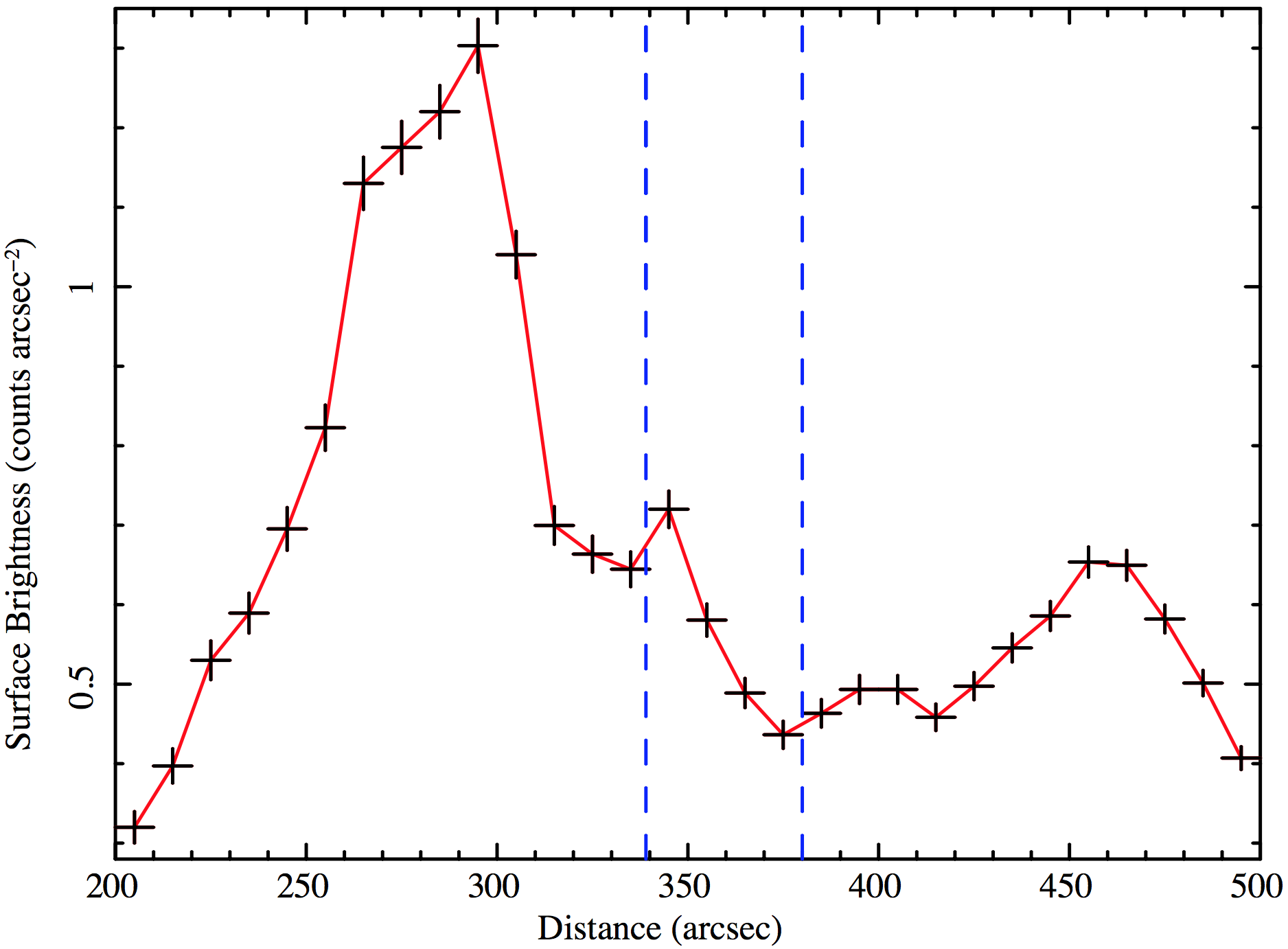}
            \includegraphics*[width=8.5cm]{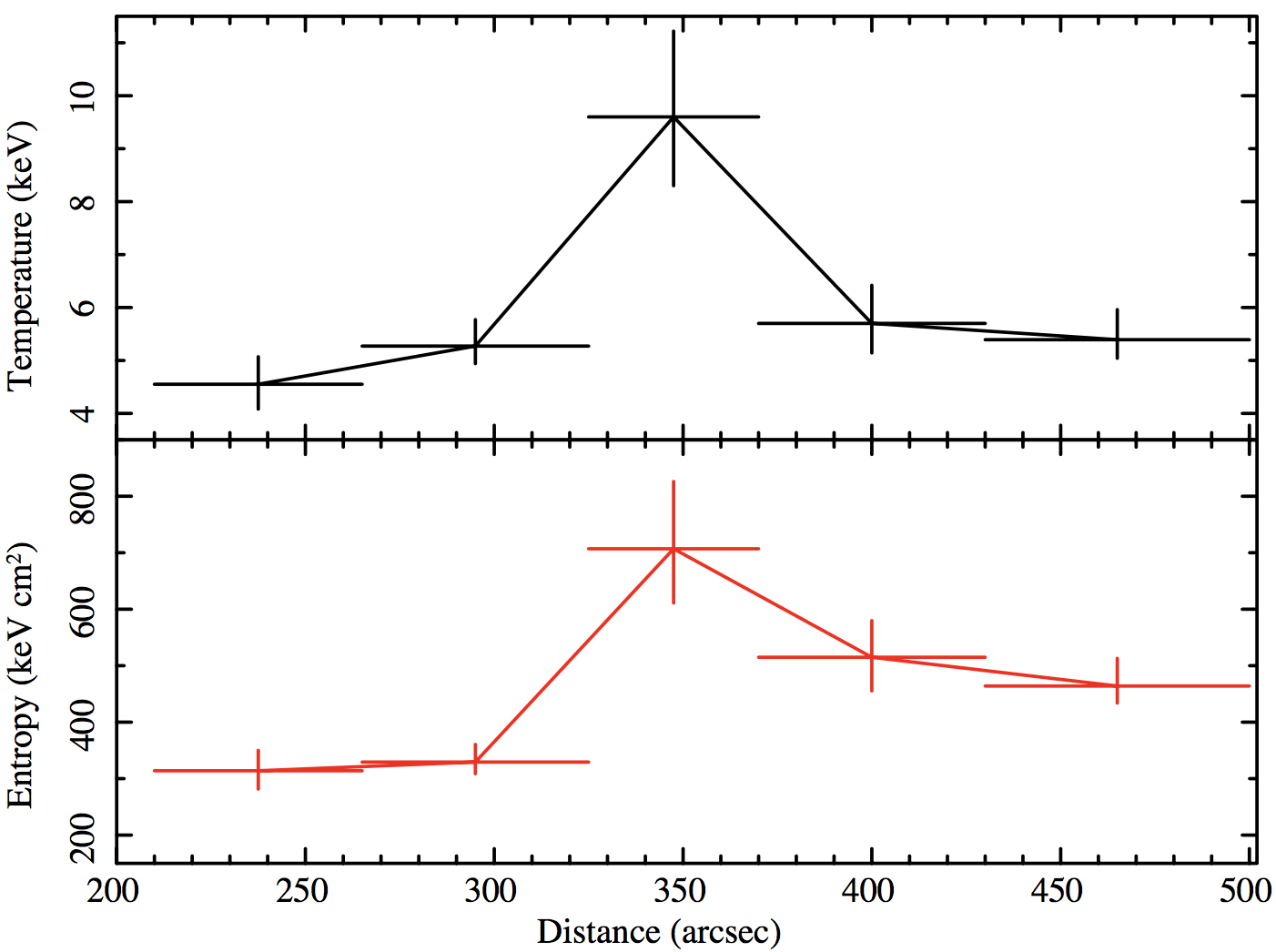}
 \caption{\textbf{Top-Left:}  All band (0.5 - 10 keV) \textit{XMM-Newton} image with the region selection, which is used to obtain surface brightness and radial profiles. \textbf{Top-Right:} The surface brightness profile in the energy range 0.5 - 2.0 keV through the  hot region, which is generated from \textit{XMM-Newton} data. \textbf{Bottom-Left:} The surface brightness profile in the energy range 0.5 - 10.0 keV through the hot region, which is generated from \textit{XMM-Newton} data. The dashed blue vertical lines represent the edges of hot region between sub-clusters. \textbf{Bottom-Right:} A demonstration of temperature radial profile across the hot region, which is generated from \textit{XMM-Newton} data.}
 \label{sur-bri} 
\end{center} 
\end{figure*}

\subsection{X-ray Morphological Parameters}
X-ray morphological parameters are very effective tools to understand the degree of disturbance for the galaxy clusters. The first parameter is the emission centroid shift, which provides useful information about  the disturbance of the gas within galaxy clusters. In a dynamically undisturbed galaxy cluster, we expect a roughly spherical symmetry; and thus, a coincidence between X-ray peak and the centroid peak. We apply the centroid shift approximation to A0141, which appears to be a merging binary galaxy cluster. The method of \citet{Mohr} is used to obtain centroid shift, which can be calculated by the following equation:

\begin{equation}
w = {\Bigg[ \frac{1}{N-1} \sum (\Delta_i - \langle \Delta \rangle)^2 \Bigg]}^{1/2} \times \frac{1}{R_{ap}},
\end{equation}
where $N$ is the number of the total aperture, $\Delta$ is the distance between X-ray peak and the centroid  of $i$th aperture, and R$_{ap}$ = 500 kpc, which is decreased in steps of 5\%. 
The second parameter is X-ray concentration, which discriminates cool core (CC) and non-cool core (non-CC) galaxy clusters. X-ray concentration parameter is defined by \citet{Santos}:

\begin{equation}
c = \frac{S(r < 100 kpc)}{S((r < 500 kpc)},
\end{equation}
where S is the surface brightness. We estimate the X-ray concentration and centroid shift parameters for both of our sub-clusters, which are presented in Table \ref{morpo}. Due to our results, we identify both A0141N and A0141S as moderately disturbed, non-cool core sub-clusters. We note that centroid shift method is less sensitive to the presence of sub-structures along the line of sight; however, the X-ray concentration parameter is not affected by this projection effect \citep{Cassano}. 
\begin{table}
\begin{center}
\caption{The X-ray morphological parameters of A0141 with $^1$\textit{XMM-Newton}  and $^2$\textit{Chandra} data. \textbf{Note:} MD = Moderately Disturbed, NCC = Non-cool core. }
\begin{tabular}{cccc} 
\hline
\hline
Cluster			& 		w					&			 c  	& Note \\	
\hline
A0141N$^{1}$	 		& 		0.034$\pm{0.004}$			&  0.113$\pm{0.004}$ & MD-NCC \\
A0141N $^{2}$			& 		0.030$\pm{0.003}$			&  0.111$\pm{0.005}$ & MD-NCC \\
A0141S $^{1}$			& 		0.039$\pm{0.004}$			&  0.104$\pm{0.005}$ & MD-NCC \\
A0141S$^{2}$			& 		0.033$\pm{0.003}$			&  0.103$\pm{0.006}$ & MD-NCC \\
\hline
\end{tabular}
\label{morpo}
\end{center}
\end{table}

\subsection{Mass Calculations}
The total dynamical X-ray mass of a galaxy cluster can be determined by density and temperature of the gas. To estimate dynamical mass, we assume an ICM in hydrostatic equilibrium and isothermal spherical symmetry. Using $\beta$ model parameters, the total dynamical X-ray mass results in \citep{LimaNeto}:
\begin{equation}
    M(r) =  \frac{3 k T_{0} \beta r_{c}}{G \mu m_{p}} \times \left(\frac{r}{r_{c}}\right)^3 \times \left(1 + \left[ \frac{r}{r_{c}} \right]^{2} \right)^{-1}    M_{\odot}.
\end{equation}

Assuming an isothermal profile for galaxy clusters, r$_{\Delta}$ is given by \citet{LimaNeto}:
\begin{equation}
 r_{\Delta} = r_{c} \left( \frac{2.3 \times 10^8 \beta <kT>}{\Delta h_{70}^2 f^2(z, \Omega_{M}, \Omega_{\Lambda}) \mu r_{c}^2} \right), 
\end{equation}
where $\beta$ is the shape parameter, r$_{c}$ is the core radius given in kpc, <kT> is the mean cluster temperature given in keV, and f$^{2}$(z, $\Omega$$_{M}$, $\Omega$$_{\Lambda}$) is the redshift evolution of the Hubble parameter.
We present the estimated masses in Table $\ref{surbri}$.

\begin{table*}
\caption{\label{beta} The best-fitting parameters of \textit{XMM-Newton} data for $\beta$-model and $r_{200}$.}
\begin{tabular}{ccccccccc} 
\hline
\hline
Region			&     r$_{c}$	& 	$\beta$  	&	 r$_{2500}$	&	 r$_{500}$  &	 r$_{200}$  &  M$_{2500}$ & M$_{500}$  & M$_{200}$ 	 \\	
				& 	(kpc) 	&		& 	(kpc)		& 	(kpc)  &  	(kpc)  &  10$^{14}$ M$_{\odot}$ & 10$^{14}$ M$_{\odot}$ & 10$^{14}$ M$_{\odot}$ 	\\
\hline
A0141N	 	&	194.6$\pm{8.97}$	&	0.56$\pm{0.01}$ & 530.8$\pm{16.0}$  &1186.8$\pm{36.4}$  & 1876.5$\pm{56.5}$ & 1.53$\pm{0.1}$ & 3.79$\pm{0.3}$  & 6.09$\pm{0.5}$ 	 \\
A0141S 		& 	189.4$\pm{12.1}$   	&	0.51$\pm{0.01}$ & 509.4$\pm{17.97}$ &1139.1$\pm{40.2}$  & 1801.1$\pm{63.5}$ & 1.35$\pm{0.1}$ & 3.35$\pm{0.3}$  & 5.38$\pm{0.5}$ 	\\
\hline
\end{tabular}
\label{surbri}
\end{table*}

%%%%%%%%%%%%%%%%%%%%%%%%%%%%%%%%%%%%%%%%%%%%%%%%%%
%%%%%%%%%%%%%%%%%%%%%%%%%%%%%%%%%%%%%%%%%%%%%%%%%%
%%%%%%%%%%				 DISCUSSION	         %%%%%%%%%%%%%%%%%%%%%
%%%%%%%%%%%%%%%%%%%%%%%%%%%%%%%%%%%%%%%%%%%%%%%%%%
%%%%%%%%%%%%%%%%%%%%%%%%%%%%%%%%%%%%%%%%%%%%%%%%%%
  \begin{figure}
 \begin{center}
  \includegraphics*[width=8.4cm]{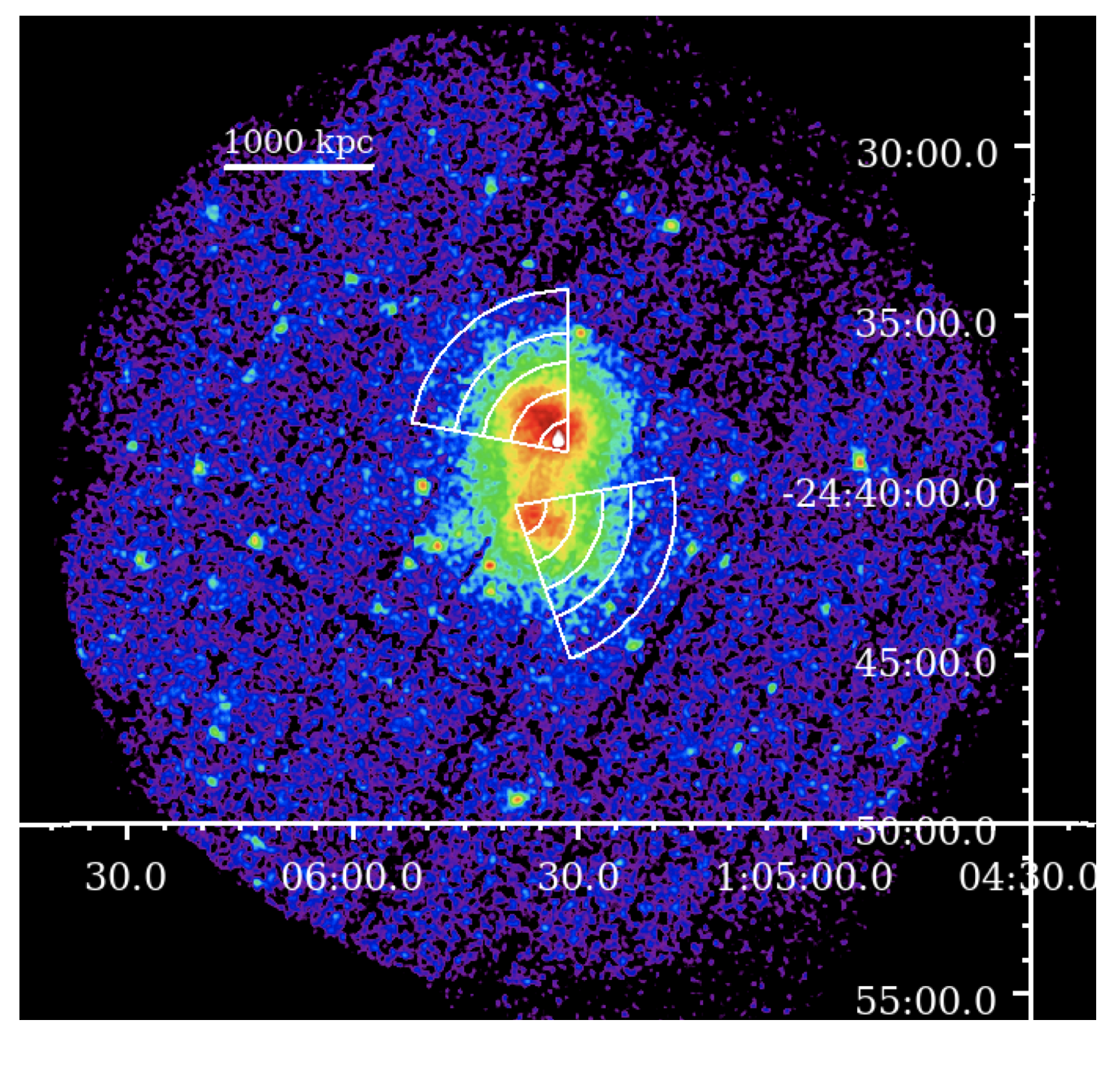}
    \includegraphics*[width=8.5cm]{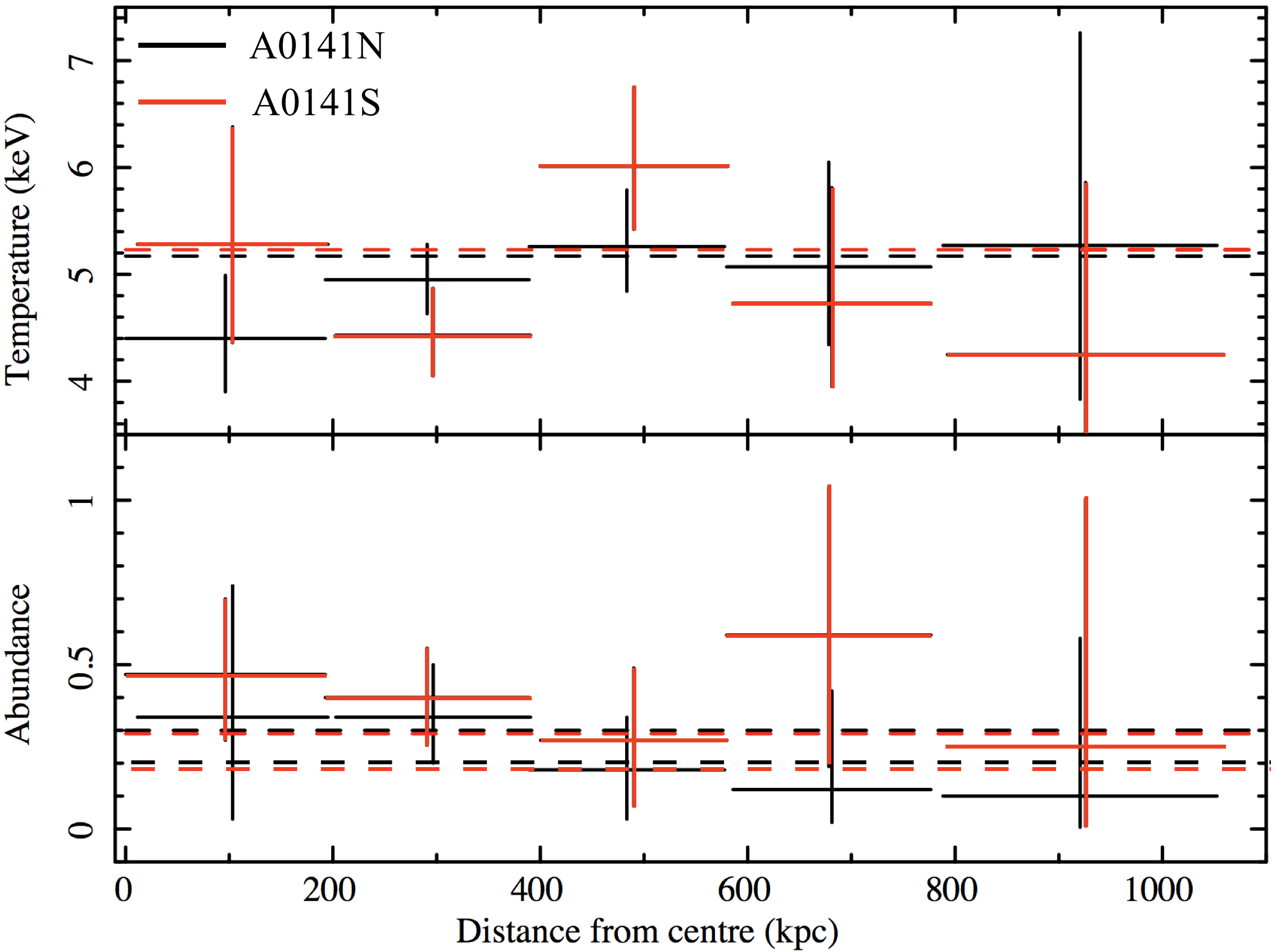}
        \includegraphics*[width=8.55cm]{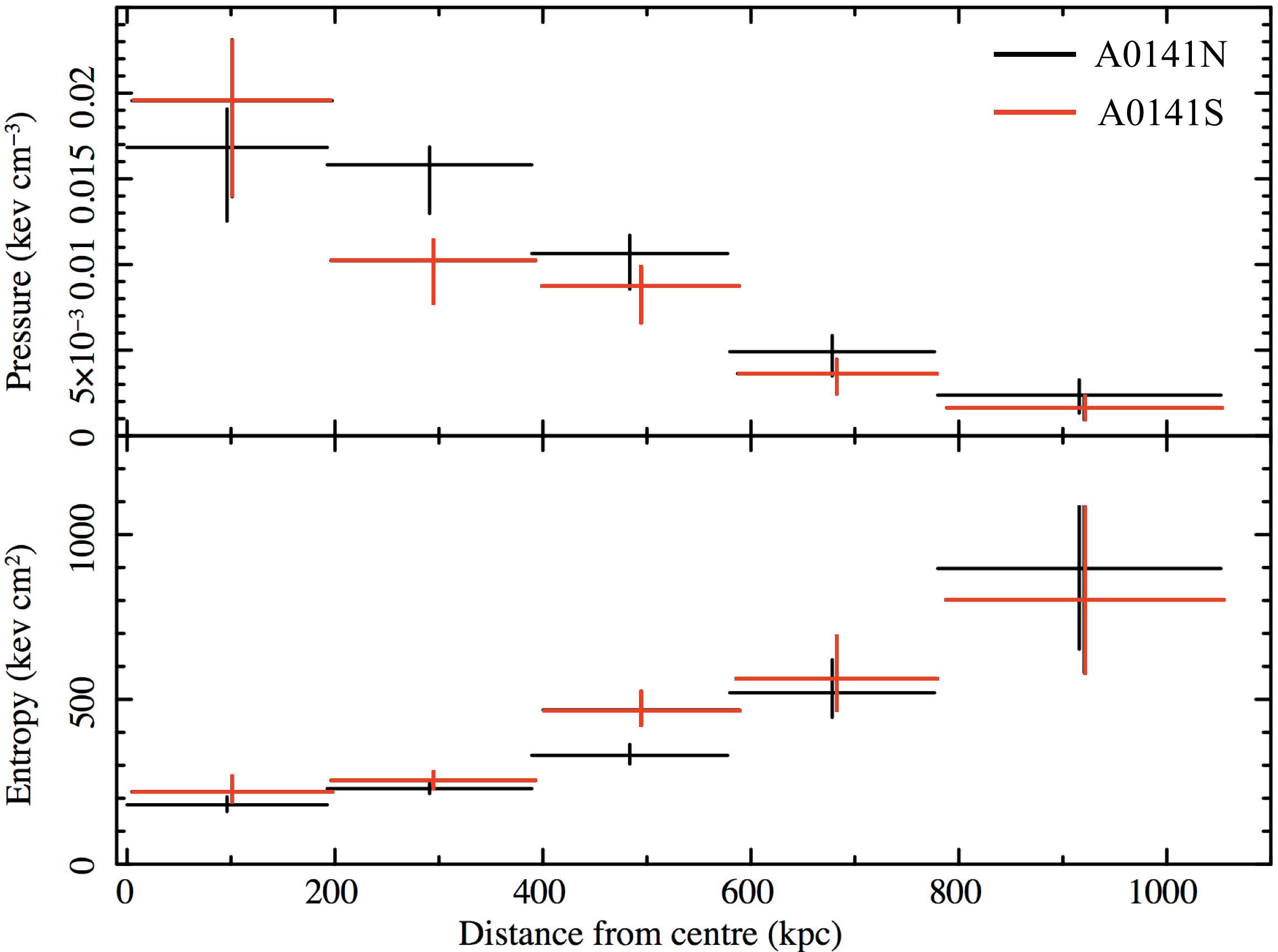}
 \caption{\label{radial} \textbf{Top:} X-ray Image of A0141, which demonstrates our region selection for radial profiles. \textbf{Middle:} The temperature and abundance radial profiles of A0141, which is obtained by \textit{XMM-Newton} data. The black and red dashed lines represent the cluster mean for A0141N and A0141S, respectively. \textbf{Bottom:} The entropy and pressure radial profiles of A0141, which is obtained by \textit{XMM-Newton} data. The black data represents the radial profile of A0141N and the red data represents the radial profile of A0141S. The specific entropies are calculated from $K$ = $kT$/$n^{2/3}_{e}$ and the specific pressures are obtained from $P$ = $kT$$n_{e}$, where $kT$ is plasma temperature and $n_{e}$ is electron density.}
\end{center} 
\end{figure}

 \section{Discussion}

\subsection{Sub-cluster structure}
Optical density and X-ray surface brightness maps reveal a bimodal galaxy cluster: A0141 with an elongation along the north-south direction (see Fig. \ref{back-sub-im} and Fig. \ref{optical-density}). A0141N is found to be the main cluster, since it is  brighter, denser and more massive on X-rays (see Figs \ref{sur-bri}, Table \ref{surbri}). A projected X-ray size of A0141 extends to 2.7 Mpc, which appears to be the edge of ICM emission. The surface brightness distribution shows two central peak points and a slight gas brightness jump in the bridge region, which is probably caused by interaction between sub-clusters. However, there is still an absence of strong X-ray emission in the bridge regions (see Fig. \ref{sur-bri}). If a close core passage occurred before or the case of a head-on merger t > 2 Gyr, we would expect compressed strong X-ray gas within sub-clusters. We note that the surface brightness profiles of both sub-clusters are fitted with a single beta profile, which is expected from non-cool core clusters. By studying the thermal structure of A0141, we find that A0141 is characterized by two cooler sub-structures ($\sim$ 4 - 6 keV) and a hotter region ($\sim$ 10 keV) between them. The central region of the main sub-cluster is colder than cluster mean, which implies that the merger interaction is still not sufficient to destroy A0141N's cool centre or the core passage has not yet destroyed the sub-cluster's cool centre. We find a hotspot ($\sim$ 10 keV) between sub-clusters; therefore, we assume that the hotspot is shock-heated by interactions between sub-clusters. We obtain a Mach number $\textit{M}$ = 1.69$^{+0.40}_{-0.37}$  from the temperature jump for shock region. If we assume that A0141S moves towards A0141N with the shock velocity V$_{infall}$ =  2081$^{+671}_{-601}$ km s$^{-1}$, the cores of sub-clusters will meet in 0.23 - 0.42 Gyr. On the other hand, surface brightness discontinuities are proposed to be related to shocks \citep{Markevitch2000}; nevertheless, they should be supported by temperature change across the edge in the opposite direction. However, we did not detect significant surface brightness discontinuities through the shock region or a sharp drop of surface brightness (see Fig. \ref{sur-bri}). We still note that such surface brightness discontinuities and sharp drops observed only in galaxy clusters with very strong shocks $M$ $\sim$ 3 \citep{Markevitch2002,Botteon,Dasadia}. In this case, approaching clusters are the main driver of the hot region in between sub-cluster. Since we did not find direct evidence for shock heating, we also suggest that the ICM can be heated by viscous dissipation. By considering tails of sub-clusters, A0141N is moving towards the south-west direction, and A0141S is moving towards the north-east direction (see Fig. \ref{back-sub-im}). The orientation of X-ray isophotes suggest that this is an off-axis merger. Furthermore, bow-like hot gas can be seen in Fig. \ref{tempmap}, which is an expected feature for off-axis mergers \citep{Takizawa}. To understand disturbance level of A0141, we calculate X-ray morphological parameters for each sub-clusters. We find centroid shifts w $\sim$ 0.03 for A0141N and w $\sim$ 0.04 for A0141S, which imply that both sub-clusters are moderately disturbed \citep{Cassano}. X-ray concentration parameters are found to be  c $\sim$ 0.11 and c $\sim$ 0.10 for A0141N and A0141S, respectively.  By comparing our results with the results by \citet{Cassano}, we identify both sub-clusters as non-cool core disturbed systems. However, the centre of A0141N has lower temperature than cluster's mean. We assume that the cool centre of A0141N has survived from the effects of mergers so far and may be destroyed in the future. Since the entropy of a system can provide useful information about merger history, we obtain entropy profiles of both systems. The entropy radial profile demonstrates that the centres of both sub-clusters have slightly low entropies (see Fig. $\ref{radial}$). The sub-clusters' core entropies are found to be  $K_{0}$ = 179.88$^{+24.13}_{-20.45}$ keV cm$^{2}$ and $K_{0}$ = 229.15$^{+47.74}_{-39.94}$ keV cm$^{2}$ for A0141N and A0141S, respectively. The resulting entropies of both sub-clusters are consistent for non cool core galaxy clusters with $kT$ $\sim$ 4 - 6 keV, which are expected to be $K_{0}$ $>$ 100 keV cm$^{2}$ \citep[e.g.][]{RoMo,McDonald}. It is known that merging events destroy cool cores of galaxy clusters and increase temperature and entropy of the cores \citep[e.g.][]{Andrade,Zuhone}. Due to temperature, X-ray morphological parameters and entropy results, we did not find a direct evidence of the merger in the cores of sub-clusters. The pressure and entropy profiles demonstrate a continuous tendency through the opposite direction of shocks; therefore, the gas within opposite direction of shocks are not disturbed (see Fig. \ref{radial}). On the other hand, the gas between sub-clusters is highly disturbed, which implies a 1:0.9 mass off-axis merger. Therefore, we compare our results with 1:1 off-axis merger simulations, which are reported by \citet{Schindler,Zuhonephd,Zuhone}. We obtain the entropy of shock-heated gas S $\sim$ 700 keV cm$^{2}$, which is consistent with our expectations for an off-axis  early stage merger system (see Fig. \ref{sur-bri}). In off-axis merger case, an equal mass approaching system can generate flattened low entropy structure between sub-clusters in t $\approx$ 1.6 Gyr \citep{Zuhone}; in this case, we find an evidence that A0141 system is in an early stage of merger t $<$ 1.6 Gyr. Due to these results, we would expect a collision between sub-clusters in t < 1.0 Gyr. After the core collision phase, we assume that the entropy of shock eventually will drop K $\sim$ 500 keV cm$^{2}$ \citep{Zuhone}. Finally, we suggest that we are witnessing an earlier phase of close core passage between sub-clusters. We note that our results are calculated as projected, and deprojected values can give better knowledge about the merger history of A0141. 

\subsection{The shift between X-ray and optical centroids}
Abell 141 is dominated by four elliptical galaxies rather than a cD galaxy \citep{Dahle,Krick}. Unfortunately, the properties of the member galaxies are not available in the literature. Therefore, the brightest central galaxy (BCG): 2MASX J01053543-243747 is identified by The Two Micron All Sky Survey \citep{Skrutskie}. We apply a 4 $\sigma$ detection likelihood threshold; however, we did not detect any X-ray counterpart for this galaxy. We assume that the faint X-ray counterpart of BCG is buried inside the ICM, which makes this source undetectable. On the other hand, the brightest X-ray peak of A0141 system coincidences with the X-ray point source: XMMU J010532.7-243844 (2MASX J01053279-2438476). A very high coincidence probability between X-ray to optical centroids ( $\sim$ \% 95 with $\sim$ 8 kpc shift) is obtained by the method explained in \citet{CAGLAR2}. Unfortunately, there is no spectroscopic redshift information for this source in the literature. By assuming this galaxy as a cluster member, we obtain an X-ray versus flux ratio X/O = -0.76, hardness radio HR (2.0 - 10.0 keV / 0.5 - 2.0 keV) = -0.40$\pm{0.24}$ and X-ray luminosity log L$_{X}$ (2.0 - 10.0 keV) = 43.46 erg s$^{-1}$. X-ray AGNs tend to have  log L$_{X}$ > 42 erg s$^{-1}$ HR > -0.55 and \citep{CAGLAR2} and X/O > -1 \citep{Fiore}. Assuming the cluster's redshift, we identify this source as an AGN due to high X/O, obscurity and X-ray luminosity. By applying B-R/R methods for cluster membership, which is suggested by \citet{Krick}, we obtain B-R = 1.82 for this source, and this result roughly makes this galaxy as a cluster member. We also apply same method for the BCG for comparison, and we obtain B-R = 2.06, which also makes BCG as cluster member expectedly. Finally, a low offset between X-ray and optical centroids of this source ($\sim$ 8 kpc) gives an impression of early stage merger system; however, we still note that a detailed optical spectroscopic study is required for this assumption.

\subsection{The radio halo}
It is known that there is a strong connection between diffuse radio emission and cluster mergers \citep[e.g.,][]{Tribble,Shimwell}. A0141 was investigated for the radio emission on different frequencies (610 MHz, 1.400 GHz) by \citet{Venturi1,Venturi2,Enblin}; however, no radio halo was detected. Surprisingly, a recent study of A0141 has proposed a discovery of an extended radio halo ($\sim$ 1.25 Mpc) at 168 MHz \citep{Duchesne}. The centre of radio halo is located in the main cluster and also has an elongation along the north-south axis. A flux density of S$_{168}$ = 110 $\pm{11}$ mJy with a large angular scale 1250 kpc and spectral index $\alpha$ $\leq$ -2.1$\pm{0.1}$ are obtained by \citet{Duchesne}, which makes this halo one of the steepest haloes known. A diffuse radio halo is quite unusual for early-stage merger systems; besides, a few early stage merging galaxy clusters are found to host radio haloes \citep[e.g.,][]{Murgia,Ogrean2015}. However, we can only speculate about the origin of this radio halo. By considering high entropy plasma, the radio halo can be interpreted as a result of major merger taking place sometime in the past (> 1 Gyr ago). Therefore, we assume that the particles are re-accelerated by the turbulence \citep[e.g.,][]{Brunetti}. In this case, the presence of a radio halo implies a post-merger system; however, we do not have enough information to give this conclusion. By considering their high PSF size (2.3 arcmin) of their survey and the presence of very bright central AGN, we still note that the existence of a radio halo is suspicious for A0141 system.

 \begin{figure}
 \begin{center}
  \includegraphics*[width=8.2cm]{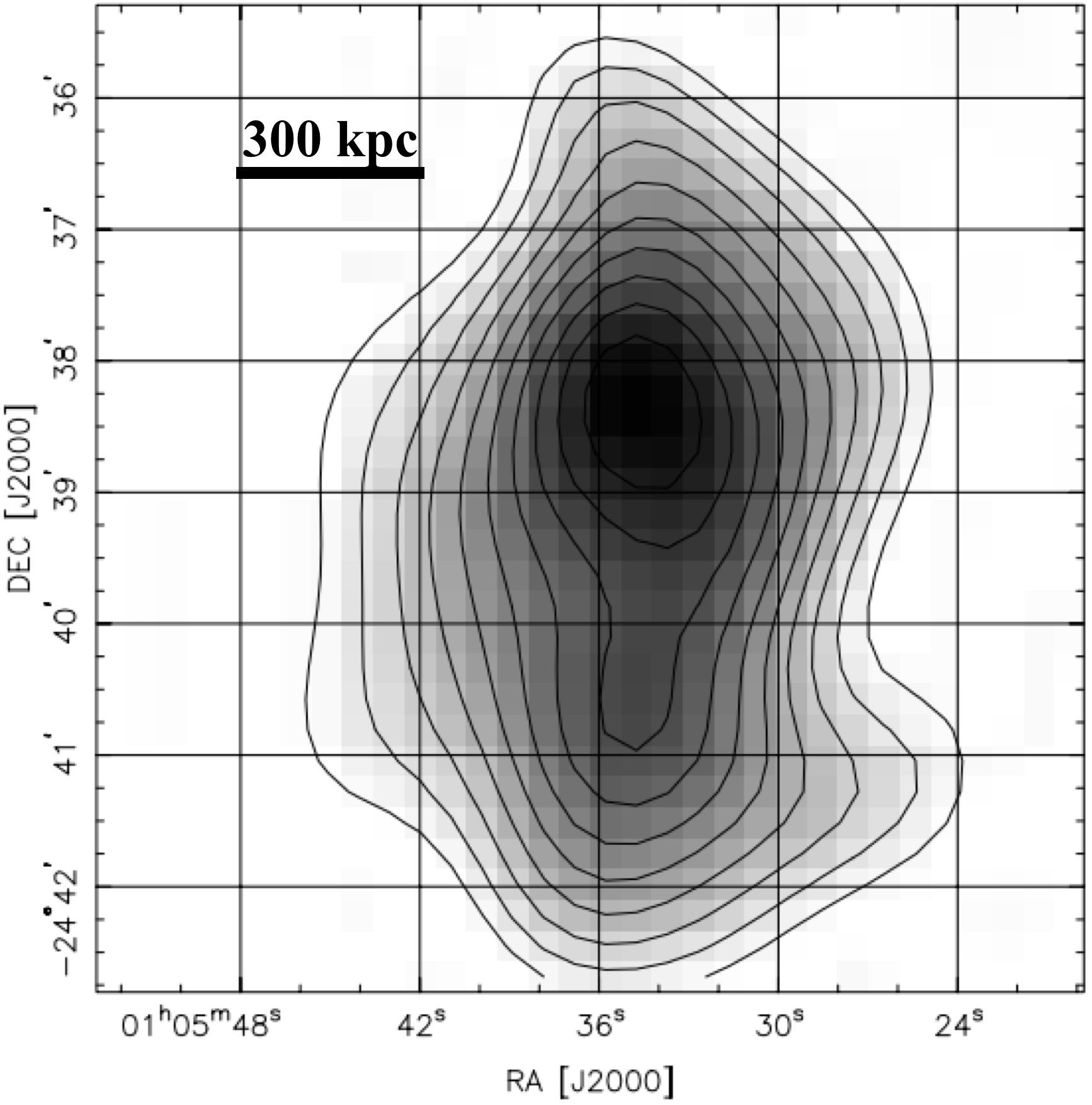}
 \caption{\label{optical-density} The galaxy number density map of Abell 141, which was reported by \citet{Dahle} using weak gravitational lensing data.}
\end{center} 
\end{figure}

%%%%%%%%%%%%%%%%%%%%%%%%%%%%%%%%%%%%%%%%%%%%%%%%%%
%%%%%%%%%%%%%%%%%%%%%%%%%%%%%%%%%%%%%%%%%%%%%%%%%%
%%%%%%%%%%				 THE END	       %%%%%%%%%%%%%%%%%%%%%
%%%%%%%%%%%%%%%%%%%%%%%%%%%%%%%%%%%%%%%%%%%% %%%%%%
%%%%%%%%%%%%%%%%%%%%%%%%%%%%%%%%%%%%%%%%%%%%%%%%%%
\section{Summary}
In this work, the structural analysis of A0141 is performed using X-ray data.The obtained features indicate that A0141 is a bimodal system, which is in an early stage of the merger. Here, we present the summary of our main findings as follows:  
\begin{itemize}

\item X-ray and optical density maps demonstrate the existence of dense region around both sub-clusters (see Fig. \ref{back-sub-im} and Fig. \ref{optical-density}). 

\item The mean temperature values of A0141N and A0141S are found to be 5.17$^{+0.20}_{-0.19}$ keV and 5.23$^{+0.24}_{-0.23}$  keV, respectively (see Table \ref{spect}). 

\item The main sub-cluster (A0141N) hosts a cooler gas in its centre (kT = 4.40 keV), which is surrounded by warmer plasma (kT = 4.95 keV) (see Fig. \ref{radial}). 

\item We find a hotspot ($\sim$ 10 keV) between sub-clusters, however, we did not find a direct evidence of shock-heating (see Table \ref{thermal}, Fig. \ref{sur-bri}).

\item A Mach number $\textit{M}$ = 1.69$^{+0.44}_{-0.39}$ is obtained from the temperature jump condition.  

\item The sub-clusters' masses are found to be M$_{200}$ = 6.09$\pm{0.5}$ $\times$ 10$^{14}$ M$_{\odot}$ and M$_{200}$ = 5.38$\pm{0.5}$ $\times$ 10$^{14}$ M$_{\odot}$ for A0141N and A0141S, respectively (see Table \ref{surbri}). 

\item The central entropies are found to be $K_{0}$ = 179.88$^{+24.13}_{-20.45}$ keV cm$^{2}$ and $K_{0}$ = 229.15$^{+47.74}_{-39.94}$ keV cm$^{2}$ for A0141N and A0141S, respectively. The resulting entropies of both sub-clusters are consistent for non-cool core galaxy clusters (see Fig. \ref{radial}).

\end{itemize}

In this work, we find some evidence that A0141 is in an earlier phase of the merger due to the following signatures; (i) the X-ray and optical centres of the brightest galaxy substantially coincidence with each other by a small displacement ($\sim$ 8 kpc), (ii) the gas within the sub-clusters' cores have not been destroyed yet, (iii) a slight brightness jump in the bridge region is detected; but, there is still an absence of strong X-ray emitting gas between sub-clusters, and (iv) there is a significantly hot region ($\sim$ 10 keV) between sub-clusters, which is generated by merger activities. We find some evidence that the system undergoes an off-axis collision; however, the cores of each sub-cluster have not been destroyed yet. Due to orientation of X-ray tails of sub-clusters, we suggest that the northern sub-cluster moves through the south-west direction, and the southern cluster moves through the north-east direction. In conclusion, we are witnessing an earlier phase of close core passage between sub-clusters. Since the optical structure of Abell 0141 is still unclear, an optical spectroscopic analysis is required. Furthermore, a longer exposure X-ray observation may constrain better statistical results. Finally, we encourage comparisons with numerical simulations in the future.    

\section*{Acknowledgement}
The author is grateful to the anonymous referee for the comments that significantly improved this article.  
The author would like to thank Silvano Molendi, Helen Russell, John Zuhone, Reinout van Weeren and Hiroki Akamatsu for their valuable comments and suggestions. The author would like to thank Haakon Dahle for optical density map and Stefano Ettori for centroid shift code. The author acknowledge financial support from the Scientific and Technological Research Council of Turkey (T\"{U}B\.{I}TAK) project number 113F117.

%%%%%%%%%%%%%%%%%%%%%%%%%%%%%%%%%%%%%%%%%%%%%%%%%%

%%%%%%%%%%%%%%%%% APPENDICES %%%%%%%%%%%%%%%%%%%%%

%%%%%%%%%%%%%%%%%%%%%%%%%%%%%%%%%%%%%%%%%%%%%%%%%%

% Don't change these lines
\bsp	% typesetting comment
\label{lastpage}
\end{document}